\documentclass[12pt]{article}
\usepackage{amsmath,amssymb,amsfonts,amsthm}

\title{Tripartite entanglement dynamics and entropic squeezing of a three-level atom interacting with a bimodal cavity field}

\author{M J Faghihi$^{1,2,3}$, M K Tavassoly$^{1,2,*}$ and M Bagheri Harouni$^{4}$ \\
 \footnotesize{$^1$ Atomic and Molecular Group, Faculty of Physics, Yazd University, Yazd, Iran} \\
 \footnotesize{$^2$ The Laboratory of Quantum Information Processing, Yazd University, Yazd, Iran} \\
 \footnotesize{$^3$ Physics and Photonics Department, Graduate University of Advanced Technology, Mahan, Kerman, Iran} \\
 \footnotesize{$^4$ Department of Physics, Faculty of Science, University of Isfahan, Hezar Jerib St. Isfahan, Iran} \\
 \footnotesize{$^*$ E-mail: mktavassoly@yazd.ac.ir}}

\begin{document}
\maketitle
%=======================================================================
%=======================================
 %=======================================

 \newcommand{\norm}[1]{\left\Vert#1\right\Vert}
 \newcommand{\abs}[1]{\left\vert#1\right\vert}
 \newcommand{\set}[1]{\left\{#1\right\}}
 \newcommand{\R}{\mathbb R}
 \newcommand{\I}{\mathbb{I}}
 \newcommand{\C}{\mathbb C}
 \newcommand{\eps}{\varepsilon}
 \newcommand{\To}{\longrightarrow}
 \newcommand{\BX}{\mathbf{B}(X)}
 \newcommand{\HH}{\mathfrak{H}}
 \newcommand{\A}{\mathcal{A}}
 \newcommand{\D}{\mathcal{D}}
 \newcommand{\N}{\mathcal{N}}
 \newcommand{\x}{\mathcal{x}}
 \newcommand{\p}{\mathcal{p}}
 \newcommand{\la}{\lambda}
 \newcommand{\af}{a^{ }_F}
 \newcommand{\afd}{a^\dag_F}
 \newcommand{\afy}{a^{ }_{F^{-1}}}
 \newcommand{\afdy}{a^\dag_{F^{-1}}}
 \newcommand{\fn}{\phi^{ }_n}
 \newcommand{\HD}{\hat{\mathcal{H}}}
 \newcommand{\HDD}{\mathcal{H}}
%==================================================

 \begin{abstract}
 In this paper, we study the interaction between a $\Lambda$-type three-level atom and two quantized electromagnetic fields which are simultaneously injected in a bichromatic cavity surrounded by a Kerr medium in the presence of the field-field interaction (parametric down conversion) and detuning parameters. By applying a canonical transformation, the introduced model is reduced to a well-known form of the generalized Jaynes-Cummings model. Under particular initial conditions which may be prepared for the atom and the field, the time evolution of state vector of the entire system is analytically evaluated. Then, the dynamics of atom is studied through the evolution of the atomic population inversion. In addition, two different measures of entanglement between the tripartite system (three entities make the system: two field modes and one atom) i.e., von Neumann and linear entropy are investigated. Also, two kinds of entropic uncertainty relations, from which entropy squeezing can be obtained, are discussed. In each case, the influences of the detuning parameters and Kerr medium on the above nonclassicality features are analyzed via numerical results, in detail. It is illustrated that the amount of the above-mentioned physical phenomena can be tuned by choosing the evolved parameters, appropriately.
 \end{abstract}

 %==============================================
 %==============================================

 %==========================================================================
 \section{Introductory remarks}\label{sec-intro}
%==========================================================================
Entanglement, a purely quantum measure of correlation with no classical counterpart, is considered to be the most important resource in the quantum information science such as quantum computation and communication \cite{QCC}, quantum dense coding \cite{QDC}, quantum teleportation \cite{QTelep}, entanglement swapping \cite{ES}, sensitive measurements \cite{SM}, and quantum telecloning \cite{QTelec}.
However, there exist other nonclassical correlations (apart from the entanglement) that have attracted a lot of attention in this field of research. Quantum discord, exhibiting the basic aspect of classical bipartite states which has been first introduced by Ollivier and Zurek \cite{QD}, is a criterion, for instance, to characterize all nonclassical correlations that has been used by the authors \cite{QD-E}. Anyway, the quantum entanglement is a vital fundamental concept in quantum information processing and plays a key role within new information technologies \cite{benenti}.
Of course, it has been shown that entanglement (of a two-qubit system) may decrease abruptly (and also non-smoothly) and tends asymptotically to zero in a finite time. This phenomenon, which is a result of the presence of quantum/classical noise, is called entanglement sudden death (ESD) \cite{ESD}. Both theoretical and experimental verifications of ESD have been recently reported the in the literature \cite{verESD}. It is also valuable to mention that contrary to the occurrence of ESD, entanglement sudden birth can be suddenly appeared \cite{QSB}. Altogether, manipulating and generating the entangled states have attracted great attention. In particular, the appearance of entanglement in the interaction between light and matter in cavity quantum electrodynamics (QED), as a simple way to produce the entangled states, is of special interest. In this respect, the atom-field entangled states have been experimentally generated through interacting a single atom with a mesoscopic field in a high-Q microwave cavity \cite{Auffeves}. \\
%
%========================================
%========================================
As is well-know, the Jaynes-Cummings model (JCM) is a fully quantum mechanical and an exact model describing the atom-field interaction \cite{JCM}. This model gives a pattern to solve the interaction between a single-mode quantized electromagnetic field and a two-level atom in the rotating wave approximation (RWA).
Based on the JCM and its generalizations, it is shown that the atom-field interaction naturally leads to the quantum entangled state. Various generalizations for the modification of the JCM, by using the multi-level atoms, multi-mode fields and multi-photon transitions and so on have been proposed in the literature throughout the recent five decades. For instance, a general formalism for a $\Lambda$-type three-level atom interacting with a correlated two-mode field has been presented in \cite{atyEPJD}. A general three-level atom interacting with a bimodal cavity field has been carried out in \cite{atyLP}. The interaction between an effective two-level atom and a three-mode field (namely finite dimensional trio-coherent state), has been studied in \cite{khalekLP}. The case of two two-level atoms interacting with two photons of the field in the presence of degenerate parametric amplifier has been examined in \cite{Khalil}.
 A model for the interaction of a three-level atom in the $\Lambda$- and $V$-configuration with a two-mode field under a multi-photon process has been presented in \cite{obadaLP-L,obadaLP-V}. The interaction between a $\Lambda$-type three-level atom and a single-mode field in a cavity including Kerr medium with intensity-dependent coupling in the presence of the detuning parameters has been studied by us \cite{us}. More recently, the nonlinear interaction between a three-level atom (in the $\Lambda$-configuration) and a two-mode cavity field in the presence of a Kerr medium and its deformed counterpart \cite{newhonarasa} has been investigated in \cite{usJOSA,usJOSA2}. In addition, a model for a moving three-level atom interacting with a single-mode quantized field with intensity-dependent coupling has been proposed in \cite{newJPB}.
In all above-mentioned studies which are only a few of the papers concerning with the JCM and its extensions, in particular, the entanglement via von Neumann entropy (as well as some of the other nonclassicality features) is calculated and discussed. \\
%
%========================================
%========================================
%
In another perspective of this field of research and in direct relation to the present work, one may consider two coupled quantized fields jointly enter within a high-Q two-mode bichromatic cavity, as is considered in \cite{abd-AOP,abdel-aty}. In this way, the atom can interact with each field individually as well as both fields. This is indeed, the main feature of these works which makes them distinguishable from previous published papers in this field. In detail,  quantum von Neumann entropy and the phase distribution function for the interaction between a two-level atom and two coupled fields simultaneously injecting in the cavity have been examined in \cite{abd-AOP}, in which there exists the field-field interaction in the non-degenerate parametric amplifier form. Also, entanglement dynamics (measured by concurrence) and coherence loss (via calculating the  linear entropy) of a tripartite system containing a two-level atom and two cavity modes in the presence of Kerr medium have been discussed in \cite{abdel-aty}. In this latter attempt, as a result of the field-field interaction, parametric down conversion has been included.\\
%
%========================================
%========================================
%
 Anyway, in this paper, we intend to outline the interaction between a three-level atom (which is assumed to be in the $\Lambda$-configuration) with two coupled quantized fields injected simultaneously in a bichromatic cavity within a centrosymmetric nonlinear (Kerr) medium in the presence of detuning parameters.
 In addition, considering the field-field interaction, parametric down conversion is properly taken into account. After considering all existing interactions appropriately in the Hamiltonian model of the entire atom-field system, it is found that, we have to apply a particular canonical transformation to reduce our complicated model to a form that can be solved analytically, based on our previous experiences on the generalized JCM \cite{us,usJOSA,usJOSA2}.
Consequently, the explicit form of the state vector of the whole system is exactly obtained by using the time-dependent Schr\"{o}dinger equation. So, briefly, the main goal of this paper is to investigate individually and simultaneously the effects of Kerr medium (containing self- and cross-action) and detuning parameters on the entanglement dynamics and some of the well-known nonclassicality features. To achieve this purpose, the time evolution of atomic population inversion is examined. Also, the amount of the DEM between subsystems (the three-level atom and the two cavity fields) by using the von Neumann entropy and the coherence loss by applying the idempotency defect are studied. In addition, number-phase and position-momentum entropic squeezing are numerically investigated, in detail. \\
%===========================================================================
More clarifying our motivations of this paper, it is valuable to give a few words on the notability of the considered physical (nonclassical) criteria, particularly entanglement and phase properties. Even though, the importance of quantum entanglement has been previously mentioned, it is instructive to point out about the entangled state due to the fields which may be utilized as an input data for new researches/observation such as in the field of quantum computation. It is shown that two electromagnetic field modes of a cavity can be applied as a universal quantum logic gate \cite{sandersPRA}. In addition, it is recently reported that entangled state can be used in quantum metrology \cite{QMet} and in the violation of Bell's inequalities \cite{VBI}. Also, regarding the number-phase entropic uncertainty relation, it should be mentioned that, studying the quantum phase properties leads to investigating the variation of the phase relation between the photons of the field. Moreover, it is demonstrated that for the single-mode JCM the evolution of the phase variance, as well as the phase distribution can carry certain information on the collapse-revival phenomenon of the corresponding atomic inversion \cite{dung}. \\
%========================================
%========================================
%
Anyway, the remainder of paper is organized as follows: In the next section, the state vector of the whole system is obtained. In section 3, in order to study the atomic dynamics, the atomic population inversion is investigated. Considering the von Neumann and linear entropies, sections 4 deals with two different regimes of entanglement. Then, based on the two kinds of entropic uncertainty relations, entropy squeezing is examined in section 5. Finally, section 6 contains a summary and concluding remarks.

%==============================================================================================================================
 \section{Theoretical formulation of the interaction model}
%==============================================================================================================================
From the standpoint of quantum mechanics, possible information during the study of any physical system is hidden in the wave function of the quantum system. This is obtained when an exact view on the existing interactions between subsystems is achieved. This section is allocated to study a three-level atom interacting simultaneously with two coupled fields in a cavity involving Kerr medium with detuning parameters. So, let us consider a model in which two quantized  electromagnetic fields oscillating with frequencies $\Omega_{1}$ and $\Omega_{2}$ interact simultaneously with a three-level $\Lambda$-type atomic system in the optical cavity which is enclosed by the Kerr medium. Also, considering the centrosymmetric nonlinear medium, self- and cross-action of the Kerr medium may be taken into account. In addition, in order to take into account the field-field interaction, parametric down conversion is properly included. It should be noted that, in the considered  atomic configuration which the levels of the atom are indicated by $|j\rangle$ with energies $\omega_{j}$, $ j=1,2,3$  (see figure \ref{scheme}), the allowed transitions are $|1\rangle\rightarrow|2\rangle$ and $|1\rangle\rightarrow|3\rangle$ whereas the transition $|2\rangle\rightarrow|3\rangle$ is forbidden in the electric-dipole approximation \cite{zubairy}. Anyway, the Hamiltonian comprising all existing interactions which describe the dynamics of the considered quantum system in the RWA can be written as ($\hbar=c=1$):
\begin{eqnarray}\label{H}
\hat{H} =\hat{H}_{A}+\hat{H}_{F}+ \hat{H}_{AF},
\end{eqnarray}
where the atomic and field parts of the Hamiltonian are given by
\begin{eqnarray}\label{H-PA}
\hat{H}_{A} &=& \sum_{j=1}^{3} \omega_{j}\hat{\sigma}_{jj}, \nonumber \\
\hat{H}_{F} &=& \sum_{j=1}^{2} \left( \Omega_{j} \hat{a}^{\dag}_{j} \hat{a}_{j} + \chi_{j} \hat{a}_{j}^{\dag 2} \hat{a}_{j}^{2} \right)
+ \chi_{12} \hat{a}_{1}^{\dag} \hat{a}_{1} \hat{a}_{2}^{\dag} \hat{a}_{2} + g_{12} \left( \hat{a}_{1}^{\dag}\hat{a}_{2} +
\hat{a}_{1}\hat{a}_{2}^{\dag}  \right),
\end{eqnarray}
and the atom-field interaction reads as
\begin{eqnarray}\label{H-PB}
\hat{H}_{AF} = \sum_{j=1}^{2} \left[ g_{1}^{(j)} (\hat{a}_{j} \; \hat{\sigma}_{12}+\hat{\sigma}_{21}\hat{a}_{j}^{\dag})
+ g_{2}^{(j)} (\hat{a}_{j}\;\hat{\sigma}_{13}+\hat{\sigma}_{31}\hat{a}_{j}^{\dag}) \right].
\end{eqnarray}
Here $\hat{\sigma}_{ij}=|i\rangle \langle j| (i,j=1,2,3)$, is the atomic ladder operator between the levels $|i\rangle$ and $|j\rangle$ and $\hat{a}_{j}$ ($\hat{a}_{j}^{\dag}$) is the bosonic annihilation (creation) operator of the field mode $j$. The constants $g_{1}^{(j)}, g_{2}^{(j)}$ determine the strength of the atom-field couplings for the mode $j$, and $g_{12}$ denotes the field-field coupling constant. In addition, $\chi_{j}$, $j = 1,2$, and $\chi_{12}$ are referred to the cubic susceptibility of the medium; $\chi_{j}$ represents the Kerr self-action for mode $j$, while $\chi_{12}$ is related to the Kerr cross-action process. \\
Now, in order to analyze the dynamics of the considered quantum system described by the Hamiltonian in equation (\ref{H}), we utilize the probability amplitudes method \cite{zubairy,us,usJOSA}. But, before using this approach, in order to simplify the complicated system by reducing the Hamiltonian introduced in relations (\ref{H})-(\ref{H-PB}) to the typical form of the JCM, we introduce the following canonical transformations
\begin{eqnarray}\label{Can-Trans}
\hat{a}_{1} = \hat{b}_{1} \cos \theta + \hat{b}_{2} \sin \theta,\;\;\;\;\;\hat{a}_{2} = \hat{b}_{2} \cos \theta - \hat{b}_{1} \sin \theta.
\end{eqnarray}
In these relations, $[\hat{a}_{i},\hat{a}_{j}^{\dag}] = [\hat{b}_{i},\hat{b}_{j}^{\dag}] = \delta_{ij}$ and the operators $\hat{b}_{i} (\hat{b}_{i}^{\dag}), i=1,2,$ have the same meaning of the operators $\hat{a}_{i} (\hat{a}_{i}^{\dag})$ while $\theta$ is the rotation angle which will be specified later. It is worthwhile to notice that under the
above transformations, the sum of the photon number of the field remains invariant, i.e., $\hat{a}_{1}^{\dag} \hat{a}_{1} + \hat{a}_{2}^{\dag} \hat{a}_{2} = \hat{b}_{1}^{\dag} \hat{b}_{1} + \hat{b}_{2}^{\dag} \hat{b}_{2}$.\\
Inserting the above transformations into the whole Hamiltonian in equation (\ref{H}) leads us to the following Hamiltonian
\begin{eqnarray}\label{H-F}
 \hat{\mathcal{H}} = \hat{H}_{0} + \hat{H}_{1},
 \end{eqnarray}
 where
 \begin{eqnarray}\label{H-0}
 \hat{H}_{0} =  \sum_{j=1}^{3} \omega_{j}\hat{\sigma}_{jj} + \sum_{j=1}^{2} \mathbf{\Omega}_{j} \hat{b}^{\dag}_{j} \hat{b}_{j},
 \end{eqnarray}
 and
 \begin{eqnarray}\label{H-1}
 \hat{H}_{1} &=& \chi \sum_{j=1}^{2} \hat{b}_{j}^{\dag 2} \hat{b}_{j}^{2}  + 2 \chi \hat{b}_{1}^{\dag} \hat{b}_{1} \hat{b}_{2}^{\dag} \hat{b}_{2} \nonumber \\
 &+& \sum_{j=1}^{2} \left[ \mu_{1}^{(j)} (\hat{b}_{j}\; \hat{\sigma}_{12}+\hat{\sigma}_{21}\hat{b}_{j}^{\dag})
 + \mu_{2}^{(j)} (\hat{b}_{j}\;\hat{\sigma}_{13}+\hat{\sigma}_{31}\hat{b}_{j}^{\dag}) \right].
 \end{eqnarray}
Paying attention to the latter relation implies the fact that, the canonical transformations in equation (\ref{Can-Trans}) preserve the invariance of the Kerr nonlinearities so long as the relation $\chi_{1} = \chi_{2} = \chi = \chi_{12}/ 2$ is satisfied. Meanwhile, we have also defined
 \begin{eqnarray}\label{H-Coeff}
 \mathbf{\Omega}_{1} &=& \Omega_{1} \cos^{2} \theta + \Omega_{2} \sin^{2} \theta - g_{12} \sin 2 \theta, \nonumber \\
 \mathbf{\Omega}_{2} &=& \Omega_{1} \sin^{2} \theta + \Omega_{2} \cos^{2} \theta + g_{12} \sin 2 \theta, \nonumber \\
 \mu_{k}^{(1)} &=& g_{k}^{(1)} \cos \theta - g_{k}^{(2)} \sin \theta, \nonumber \\
 \mu_{k}^{(2)} &=& g_{k}^{(1)} \sin \theta + g_{k}^{(2)} \cos \theta, \;\;\;  k=1,2,
 \end{eqnarray}
in which the rotation angle $\theta$ is still unknown and must be determined. For this purpose, the evanescent wave terms from the Hamiltonian related to the field and field-field interaction should be avoided. Therefore, one may set the particular choice of angle $\theta$ as
 \begin{eqnarray}\label{r}
 \theta = \frac{1}{2} \tan^{-1} \left( \frac{2  g_{12}}{ \Omega_{2} - \Omega_{1} } \right).
 \end{eqnarray}
With the above selection of $\theta$, the field-field coupling parameter $g_{12}$ will then be in the form
 \begin{eqnarray}\label{g}
  g_{12} = \frac{\delta \left( \Omega_{2} - \Omega_{1} \right) }{1 - \delta^{2}},
 \end{eqnarray}
where we have set $ g_{1}^{(1)}/g_{1}^{(2)} = \delta = g_{2}^{(1)}/g_{2}^{(2)}  $. \\
Looking deeply at the relations (\ref{H-0}) and (\ref{H-1}) and comparing them with equations (\ref{H})-(\ref{H-PB}) indicates clearly that the considered canonical transformations simplify the interaction Hamiltonian by removing the field-field interaction. In this way, our presented model is reduced to a usual form of the JCM. \\
Anyway, for next purpose, it is convenient to rewrite the Hamiltonian (\ref{H-F}) in the interaction picture. Accordingly one may arrive at
\begin{eqnarray}\label{VI}
V_{I}(t) &=& \mu \left[ \hat{b}_{2}  \hat{\sigma}_{12} \exp (- i \Delta_{2} t ) +  \gamma \; \hat{b}_{2}  \hat{\sigma}_{13} \exp (- i \Delta_{3} t )  \right] + \mathrm{h.c.}, \nonumber \\
&+& \chi \sum_{j=1}^{2} \hat{b}_{j}^{\dag 2} \hat{b}_{j}^{2}  + 2 \chi \hat{b}_{1}^{\dag} \hat{b}_{1} \hat{b}_{2}^{\dag} \hat{b}_{2},
\end{eqnarray}
 with
 \begin{eqnarray}\label{VI-Coeff}
  \mu = g \sqrt{1 + \delta^{2}},\;\;\;  \gamma = \mu_{2}^{(2)}/ \mu_{1}^{(2)}, \;\;\;  \Delta_{k} = \mathbf{\Omega}_{2} - (\omega_{1} - \omega_{k}),\;\;\;k=2,3,
 \end{eqnarray}
where, for simplicity, we have set $g_{1}^{(2)} = g$. \\
Let us consider the wave function $|\psi(t)\rangle$ corresponding to the whole system at any time $t$ to be in the form
\begin{eqnarray}\label{say}
\hspace{-2cm} |\psi(t)\rangle &=& \sum_{n_{1}=0}^{+\infty} \sum_{n_{2}=0}^{+\infty} q_{n_{1}} q_{n_{2}} \Big[ A(n_{1},n_{2},t) |1,n_{1},n_{2} \rangle
 + B(n_{1},n_{2}+1,t) |2,n_{1},n_{2}+1\rangle  \nonumber \\
 \hspace{-2cm} &+& C(n_{1},n_{2}+1,t)|3,n_{1},n_{2}+1\rangle \Big]
 \end{eqnarray}
where $q_{n_{1}}$ and $q_{n_{2}}$ are the probability amplitudes of the initial state of the radiation field of the cavity. Besides, $A,B$ and $C$ are the atomic probability amplitudes which have to be determined. Considering the probability amplitudes technique, and after some lengthy but straightforward manipulations, the probability amplitudes $A$, $B$ and $C$ (specifying the explicit form of the wave function of whole system) are given by
 \begin{eqnarray}\label{abc}
 \hspace{-2cm}A(n_{1},n_{2},t) &=& - \sum_{m=1}^{3}(\vartheta_{m} + V_{2}) \Upsilon_{m} \; e^{ i ( \vartheta_{m} -  \Delta_{2}) t }, \nonumber \\
 \hspace{-2cm}B(n_{1},n_{2}+1,t) &=& f_{1} \sum_{m=1}^{3} \Upsilon_{m} \; e^ { i \vartheta_{m} t }, \nonumber \\
 \hspace{-2cm}C(n_{1},n_{2}+1,t) &=& \frac{1}{f_{2}} \sum_{m=1}^{3}\Big[(\vartheta_{m}+V_{2})
 (\vartheta_{m} + V_{1} - \Delta_{2})-f_{1}^{2}\Big] \Upsilon_{m} \; e^ {i (\vartheta_{m} + \Delta_{3} - \Delta_{2}) t },
 \end{eqnarray}
 where
 \begin{eqnarray}\label{vkardan}
 \vartheta_{m}&=&-\frac{x_{1}}{3}+\frac{2}{3} (x_{1}^{2}-3x_{2})^{1/2} \cos\left[ \varphi +\frac{2 \pi (m-1)}{3} \right],\;\;\;\;\;m=1,2,3, \nonumber \\
 \varphi &=& \frac{1}{3}\cos^{-1}\left[ \frac{9x_{1}x_{2}-2x_{1}^{3}-27x_{3}}{2(x_{1}^{2}-3x_{2})^{3/2}}\right],
 \end{eqnarray}
 with
 \begin{eqnarray}\label{x123}
 x_{1} &=& V_{1} + 2 V_{2} + \Delta_{3} - 2 \Delta_{2}, \nonumber \\
 x_{2} &=& (V_{1} + V_{2} - \Delta_{2})(V_{2} + \Delta_{3} - \Delta_{2}) + V_{2} (V_{1} - \Delta_{2}) - f_{1}^{2} - f_{2}^{2}, \nonumber \\
 x_{3} &=& V_{2} \left[ ( V_{1} - \Delta_{2})(V_{2} + \Delta_{3} - \Delta_{2}) - f_{1}^{2} - f_{2}^{2} \right] - f_{1}^{2} (\Delta_{3} - \Delta_{2}).
 \end{eqnarray}
 Also, in the above relations we have defined
 \begin{eqnarray}\label{def}
 V_{1} &=& \chi \left[ n_{1} ( n_{1} - 1 ) + n_{2} ( n_{2} - 1 ) + 2 n_{1} n_{2} \right], \nonumber \\
 V_{2} &=& \chi \left[ n_{1} ( n_{1} - 1 ) + n_{2} ( n_{2} + 1 ) + 2 n_{1} ( n_{2} + 1 ) \right], \nonumber \\
 f_{1} &=& \mu \sqrt{ n_{2} + 1 },    \;\;\;    f_{2} = \gamma f_{1}.
 \end{eqnarray}
 Finally, the coefficients $\Upsilon_{m}$ can be obtained by determining the initial conditions for the probability amplitudes. By adopting the atom initially in the excited state, i.e. $A(0)=1$, $B(0) = 0 = C(0)$, one arrives at
 \begin{eqnarray}\label{b123}
 \Upsilon_{m}=\frac{\vartheta_{k} + \vartheta_{l} + V_{1} + V_{2} - \Delta_{2}}{\vartheta _{mk} \vartheta _{ml}},           \;\;\;\;\;\;      m \neq k \neq l = 1,2,3,
 \end{eqnarray}
 where $ \vartheta _{mk} = \vartheta_{m} - \vartheta_{k} $. Hence, the exact form of the wave function $|\psi(t)\rangle$ introduced in (\ref{say}) is explicitly derived. \\
 It is now necessary to emphasize the fact that investigating the nonclassicality features can be perform by considering arbitrary amplitudes of the initial states of the field such as number, phase, coherent or squeezed state. However, since the coherent state (the laser field far above the threshold condition \cite{zubairy}) is more accessible than other typical field states, we shall consider the fields to be initially in the coherent state
 \begin{eqnarray}\label{amplitude}
 \hspace{-2cm}|\alpha_{1}, \alpha_{2} \rangle = \sum_{n_{1} = 0}^{+\infty}  \sum_{n_{2} = 0}^{+\infty} q_{n_{1}} q_{n_{2}} |n_{1},n_{2} \rangle, \;\;\;\;q_{n_{i}} = \exp \left( -\frac{ |\alpha_{i}|^{2} }{2}\right) \frac{\alpha_{i}^{n_{i}}}{\sqrt{n_{i}!}},     \;\;\;\;         i=1,2,
 \end{eqnarray}
 in which $|\alpha_{i}|^{2}$, $i=1,2$, display the mean photon number (intensity of light) of mode $i$. Now, we are able to study the nonclassicality criteria of the state vector of the considered atom-field system, in which the appearance of the nonclassicality behaviour implies the fact that the state of the system is `nonclassical' with no classical analogue. We would like to end this section by emphasizing on the significance of the nonclassical states in which they have recently received considerable attention in various fields of research, such as quantum optics, quantum cryptography and quantum communication \cite{Nonclassical-Application}.
 %
 %===================================================================================================================================
 \section{Atomic population inversion}
 %====================================================================================================================================
We are now in a position to analyze the atomic dynamics, specially the time evolution of an important quantity, namely the atomic population inversion. Along studying the time evolution of this quantity, we discuss the collapse and revival phenomenon, as a consequence of quantum interference in phase space, which is originated in the discreteness nature of the photon number distribution of the initial field \cite{zubairy}. It should be mentioned that the realization of this  phenomenon has been experimentally reported in the literature \cite{ObsCR}. For the present model which contains a `three-level' $\Lambda$-type system, the atomic inversion is introduced as the difference between the excited state (level $|1\rangle$) and the sum of two lower states (levels $|2\rangle$ and $|3\rangle$) probabilities. Consequently, this may be defined as follows \cite{atyLP,Inv}:
 \begin{eqnarray}\label{inversion}
W(t) = \rho_{11}(t) - (\rho_{22}(t) + \rho_{33}(t)),
 \end{eqnarray}
 in which the matrix elements of atomic density operator are generally given by
 \begin{eqnarray}\label{rhoshekl}
 \rho_{i j}(t) = \sum_{n=0}^{+\infty} \sum_{m=0}^{+\infty} \langle n,m, i | \psi(t) \rangle \langle \psi(t) | n,m, j \rangle,\;\;\; i ,  j = 1,2,3.
 \end{eqnarray}
Figure \ref{Finversion} shows the evolution of the atomic population inversion against the scaled time $\tau = g t$ for initial mean number of photons fixed at $|\alpha_{1}|^{2} = 10=|\alpha_{2}|^{2}$. The left (right) plots of this figure correspond to $\gamma = 1$ ($\gamma = 2$) where $\gamma$ denotes the ratio of
atom-field couplings. In figure \ref{Finversion}(a), the exact resonant case is assumed ($\Delta_{2} = \Delta_{3} =0$) whereas the Kerr medium is absent ($\chi = 0$). Figure \ref{Finversion}(b) is plotted to indicate the effect of detuning parameters ($\Delta_{2}/g =7, \; \Delta_{3}/g = 15$), while figure \ref{Finversion}(c)
concentrates on the effect of Kerr medium ($\chi /g = 0.4$). The effect of detuning parameters together with the Kerr nonlinearities is depicted in figure \ref{Finversion}(d). \\
In detail, it is seen from both plots of figure \ref{Finversion}(a) that, the atomic inversion oscillates between some minimum and maximum values. Also, the typical collapse and revival, as the pure quantum mechanical phenomenon, can be observed. The temporal behaviour of population inversion is changed when the effect of detuning parameters is regarded (figure \ref{Finversion}(b)), in which this quantity periodically varies between the negative and positive values (the atomic inversion is observed, but not complete).
Looking at the left plot of figure \ref{Finversion}(c) indicates that the atomic inversion varies in the negative region near its minimum value at all times. While from the right one, it is observed that the amount of this quantity moves toward the positive region as time elapses.
Figure \ref{Finversion}(d) which studies simultaneously the effects of detuning and Kerr medium shows that the population inversion gets negative values close to its minimum value at all times. \\
Finally, it is found that, in the presence of the detuning parameters, periodic behaviour together with the maximally inversion is clearly observed. Also, this quantity gets negative values at all times when the Kerr medium is entered. In addition, changing the ratio of atom-field couplings has
no effective role on revealing the population inversion. At last, in the absence of all nonlinear couplers, the typical collapse and revival exhibition as a nonclassical phenomenon can be observed.
 %
 %===================================================================================================================================
 \section{Quantum entanglement}
 %===================================================================================================================================
In this section, we are going to survey the entanglement phenomenon using von Neumann entropy as a measure of the DEM and linear entropy which sometimes known as idempotency defect from which the coherence loss (the degree of mixture of the state) can be investigated.
 %===================================================================================================================================
 \subsection{Von Neumann entropy and DEM}
 %===================================================================================================================================
Quantum correlation between spatially separated objects of a system leads to an intriguing feature known as entanglement. To establish the significance of this quantity, it is enough to state that many interesting applications of quantum computation and quantum information are concluded from the concept of entanglement. Also, this is one of the main parts for the execution of quantum information processing devices \cite{qipd}. Before proceeding, it is instructive to clarify the concept of entangled and unentangled states. A pure state of bipartite quantum system (two-particle or two-part system) is separable (unentangled) if and only if the reduced density operators represent pure states. Since entanglement means as the absence of separability, a pure state is entangled if and only if the reduced density operators for the subsystems describe mixed states \cite{Garrison}, that is, quantum state cannot be described by a direct product of the quantum states of the subsystems. In order to understand the DEM for the pure state of the considered system (atom and field), the reduced quantum entropy or von Neumann entropy satisfying the general conditions consist of Schmidt decomposition, local invariance, continuity and additivity, is a good measure of entanglement \cite{chuang}. Entropy forms the core of classical information theory ({\it Shannon's entropy}) and quantum information theory ({\it von Neumann's entropy}) which measures how much uncertainty exists in the state of a physical system. The entropy of the field is a criterion which displays the strength of entanglement; higher (lower) entropy means the greater (smaller) DEM. Before obtaining the DEM based on von Neumann reduced entropy, recalling the important theorem of Araki and Leib \cite{araki} is of special importance. According to this theorem, for any two-component quantum system (here, atom and field), the entropies should satisfy the following triangle inequality
 \begin{eqnarray}\label{valen}
 |S_{A}(t)-S_{F}(t)|\leq S_{AF}(t) \leq S_{A}(t)+S_{F}(t),
 \end{eqnarray}
where the subscripts `A' and `F' refer to the atom and the field, respectively and $S_{AF}$ denotes the total entropy of the atom-field system. This theorem implies that, if at the initial time, the field and the atom are in pure states, the total entropy of the system is zero and remains constant. As a result, if the system is initially prepared in a pure state, at any time $t>0$, the entropy of the field is equal to the atomic entropy \cite{phoenix}. Therefore, both atomic and field entropies are suitable measures of the entanglement. In the following, we concentrate on the evolution of the atomic entropy against time to obtain the DEM. According to the von Neumann entropy, as an entanglement criterion, the entropy of the atom and the field are defined through the corresponding reduced density operator by
 \begin{eqnarray}\label{ventd}
 S_{A(F)}(t)=-\mathrm{Tr}_{A(F)} \left(\hat{\rho}_{A(F)}(t) \ln \hat{\rho}_{A(F)}(t) \right),
 \end{eqnarray}
where $\hat{\rho}_{A(F)}(t)$ is the reduced density matrix of the atom (field). \\
Following the procedure of \cite{usJOSA,newJPB}, we can express the entropy of the field/atom by the following relation \cite{pk3}
 \begin{eqnarray}\label{sff}
 \mathrm{DEM}(t) = S_{F}(t) = S_{A}(t) = - \sum_{j=1}^{3}\zeta_{j} \ln \zeta_{j},
 \end{eqnarray}
where $\zeta_{j}$, the eigenvalues of the reduced atomic density matrix, are given by Cardano's method as \cite{kardan}
 \begin{eqnarray}\label{ventkardan}
 \zeta_{j}&=&-\frac{1}{3}\xi_{1}+\frac{2}{3}\sqrt{\xi_{1}^{2}-3\xi_{2}}\cos\left[\varrho+\frac{2}{3}(j-1)\pi \right],\;\;\;j=1,2,3, \nonumber \\
 \varrho &=& \frac{1}{3}\cos^{-1}\left[ \frac{9\xi_{1}\xi_{2}-2\xi_{1}^{3}-27\xi_{3}}{2(\xi_{1}^{2}-3\xi_{2})^{3/2}}\right],
 \end{eqnarray}
with
 \begin{eqnarray}\label{vzal}
 \hspace{-2cm}\xi_{1} &=& -\rho_{11}-\rho_{22}-\rho_{33}, \nonumber \\
 \hspace{-2cm}\xi_{2} &=& \rho_{11}\rho_{22}+\rho_{22}\rho_{33}+\rho_{33}\rho_{11} -\rho_{12}\rho_{21}-\rho_{23}\rho_{32}-\rho_{31}\rho_{13}, \nonumber \\
 \hspace{-2cm}\xi_{3} &=& -\rho_{11}\rho_{22}\rho_{33}-\rho_{12}\rho_{23}\rho_{31}-\rho_{13}\rho_{32}\rho_{21} +\rho_{11}\rho_{23}\rho_{32}+\rho_{22}\rho_{31}\rho_{13}+\rho_{33}\rho_{12}\rho_{21},
 \end{eqnarray}
where the parameter $\xi_{1}$ is precisely equal to $-1$ and the matrix elements have been introduced in equation (\ref{rhoshekl}). The relation (\ref{sff}) displays the variation of the field (atomic) entropy versus time. By this equation the DEM between the atom and field is also measured, i.e., the subsystems are disentangled if the DEM in equation (\ref{sff}) tends to zero.\\
Our presented results in figure \ref{Fentanglement} indicate the time evolution of the field entropy against the scaled time $\tau$ for different chosen parameters similar to figure \ref{Finversion}, where the left (right) plots refer to the similar (different) atom-field couplings regime.
Figure \ref{Fentanglement}(a) shows that in the resonance case and in the absence of Kerr nonlinearities, the DEM has a random behavior. Considering both left and right plots of this figure, it is seen that the amount of DEM is reduced when $\gamma = 2$ (the right plot).
In figure  \ref{Fentanglement}(b), the influence of the detuning parameters is examined. Due to the presence of these parameters, the DEM between subsystems is diminished, while changing the ratio of couplings can lead to an increase of the amount of entanglement.
In figure  \ref{Fentanglement}(c) in which the effect of Kerr medium is considered, the DEM is strongly descended (left plot). By setting $\gamma = 2$, the amount of this quantity is considerably ascended when the time proceeds.
Figure \ref{Fentanglement}(d) demonstrates simultaneously the effects of Kerr medium and detuning parameters in the presence of Kerr nonlinearities. This plot shows that the detuning causes an increase of the DEM. Although, replacing $\gamma = 1$ by $\gamma = 2$ reduces the amount of entanglement. \\
We end this section with mentioning that, the existence of nonlinearities (Kerr medium and detuning parameters) diminishes the amount of entanglement (in the similar coupling constants). Also, it can be stated that in the presence of Kerr medium, entanglement between subsystems is obviously destroyed. It should be mentioned that, changing the ratio of atom-field couplings possesses remarkable role on the amount of the DEM. In detail, in the presence/absence of all nonlinearities the DEM is decreased while in the presence of either detuning parameters or Kerr medium, the amount of entanglement is increased.

 %===================================================================================================================================
 \subsection{Linear entropy and coherence loss}
 %===================================================================================================================================
Idempotency defect, a simple and direct measure of the degree of mixture of the state of a system, is a useful quantity to understand the degree of decoherence (coherence loss) \cite{abdel-aty,idempotency1}. This quantity, which can be regarded as the lowest order approximation to the von Neumann entropy, is a good criterion to understand the purity loss of the quantum system. This parameter which is measured by the linear entropy has been introduced as \cite{idempotency2}
 \begin{eqnarray}\label{LinEnt}
 S (\hat{\rho}) = \mathrm{Tr} \left[ \hat{\rho}_{A}(t) (1 - \hat{\rho}_{A}(t))  \right],
 \end{eqnarray}
where $\hat{\rho}_{A}(t)$ is the atomic density operator. The latter relation indicates that, the linear entropy is zero for a pure state, i.e., $\hat{\rho}_{A}(t) = \hat{\rho}_{A}^{2}(t)$. Consequently, the nonzero values of this indicator show the non-purity of the state of the system. Also, it may be noted that, maximally entanglement as well as the most mixed state is appeared whenever idempotency defect gets the value $1$. \\
In figure \ref{Floss} we have plotted the idempotency defect, from which the coherence loss is studied, in terms of scaled time $\tau$ for different chosen parameters assumed in figure \ref{Finversion}. In plotting this figure, the left (right) plots again correspond to the case $\gamma = 1$ ($\gamma = 2$).
The left plots of figures \ref{Floss}(a) and \ref{Floss}(b) indicate that, the detuning parameters reduces the coherence, when the Kerr medium is disregarded. Figure  \ref{Floss}(c) studies the effect of Kerr medium. In this case, it is seen that the degree of mixture of the state is rapidly attenuated. Entering the detuning parameters leads to an increase in the amount of this quantity (left plots of figure \ref{Floss}(d)).
It is valuable to pay attention to the right plots of this figure. From figure \ref{Floss}(c) (\ref{Floss}(d)) it is observed that, coherence loss can be increased (decreased) when the ratio of coupling constants gets the value $2$ ($\gamma = 2$). \\
By considering the presented results depicted in figure \ref{Floss}, one can deduce that coherence loss is attenuated due to nonlinear couplers. Also, in the presence of Kerr medium, selecting different values of coupling constants may enhance the decoherence.
%
 %===================================================================================================================================
 \section{Entropic uncertainty relations and entropy squeezing}
 %===================================================================================================================================
This section is devoted to the study of two types of nonclassicality features by using the number-phase and position-momentum entropic uncertainty relations. So, two kinds of {\it entropy squeezing} parameters are introduced and in each case, the squeezing condition of the state vector of the whole system is
investigated.
%
 %===================================================================================================================================
 \subsection{Number-phase entropic uncertainty relation and entropy squeezing}
 %===================================================================================================================================
As is well-known, studying the dynamics of the phase distribution of the photons of the field can be thought of as an analysis of the variation of the phase relation between atom and photons \cite{nakano}. Recently, quantum phase distribution and squeezing phenomenon in number/phase operators of various physical systems with known discrete spectrum $e_{n}$ \cite{en} have been reported by one of us in \cite{honar}. Also, the number-phase entropic uncertainty relation and the number-phase Wigner function of generalized coherent states associated with some solvable quantum systems which posses non-degenerate spectra are investigated \cite{honarasaphase}. \\
Now, we are going to examine the `entropy squeezing' for considered tripartite system in terms of entropies of {\it number} and {\it phase} operators. Over the years, many attempts such as Dirac \cite{Dirac} (the earliest endeavor), Susskind-Glogower \cite{SG} and Pegg-Barnett \cite{PB} approaches have been performed to introduce a formalism for describing the quantum phase. Altogether, defining a uniquely acceptable phase operator for the electromagnetic field is still an appealing and of course open problem in quantum mechanics \cite{knight}. However, since the formalism of Pegg and Barnett by introducing a Hermitian phase operator overcomes some obstacles of the Susskind-Glogower phase operator, in this section, the phase properties of the fields are evaluated by using the Pegg-Barnett approach \cite{ApPB}. According to this method, all observables corresponding to the phase properties are defined in an $(s+1)$-dimensional space, in which they constitute the eigenvalue equations with $(s+1)$  eigenstates (orthonormal phase states). Therefore, a complete set of $(s+1)$ orthonormal phase states (of a single-mode field) is defined by
  \begin{eqnarray}\label{pb1mode}
  |\theta_{p} \rangle = \frac{1}{\sqrt{s+1}}\sum_{n=0}^{s}\exp(in\theta_{p})| n \rangle,
  \end{eqnarray}
where $\{| n \rangle\}_{n=0}^{s}$ are the number states and $\theta_{p} = \theta_{0} + \frac{2\pi p}{s+1}$, $p = 0,1,...,s$, with the arbitrary value $\theta_{0}$. Note that in the calculations, the parameter $s$ tends finally to infinity. Since we are dealing with the two modes in the present atom-field interaction, it is necessary to extend the definition of orthonormal phase states. Accordingly, we propose the two-mode phase state $|\theta_{p},\theta_{q} \rangle$ which should be necessarily introduced as
  \begin{eqnarray}\label{pq2mode}
  |\theta_{p},\theta_{q} \rangle = \frac{1}{s+1}\sum_{n=0}^{s}\sum_{m=0}^{s} \exp\left(i n \theta_{p}\right) \exp\left( i  m \theta_{q}\right) | n,m \rangle,
  \end{eqnarray}
with
  \begin{eqnarray}\label{thetapq}
  \theta_{p}=\theta_{0}+\frac{2\pi p}{s+1},       \;\;     \theta_{q}=\theta_{0}+\frac{2\pi q}{s+1},       \;\;\;\;     p,q = 0,1,...,s,
  \end{eqnarray}
and $\theta_{0}$ is an arbitrary value. By the way, we are able to define the two-mode Pegg-Barnett phase distribution function
\begin{eqnarray}\label{pthetapq}
\mathcal{P}_{\phi}(\theta_{p},\theta_{q})=\lim_{s\rightarrow
+\infty} \left( \frac{s+1}{2\pi} \right)^{2} \langle
\theta_{p},\theta_{q}|\hat{\rho}_{F}|\theta_{p},\theta_{q} \rangle,
\end{eqnarray}
By inserting the relation (\ref{pq2mode}) into equation (\ref{pthetapq}) and paying attention to the state vector of the whole system proposed in equation (\ref{say}), in which all of its time-dependent coefficients have been exactly obtained, for the phase distribution function associated with the two modes of the cavity fields we arrive at
  \begin{eqnarray}\label{pthetapqfinal}
  \hspace{-2cm}\mathcal{P}_{\phi}(\theta_{1},\theta_{2}) &=& \frac{1}{4 \pi^{2}} \left| \sum_{n_{1}=0}^{+\infty}\sum_{n_{2}=0}^{+\infty} q_{n_{1}} q_{n_{2}} A(n_{1},n_{2},t) \exp(- i n_{1} \theta_{1}) \exp(- i n_{2} \theta_{2}) \right|^{2} \nonumber \\
  \hspace{-2cm}&+& \frac{1}{4 \pi^{2}} \left| \sum_{n_{1}=0}^{+\infty}\sum_{n_{2}=0}^{+\infty} q_{n_{1}} q_{n_{2}} B(n_{1},n_{2}+1,t) \exp(- i n_{1} \theta_{1}) \exp(- i n_{2} \theta_{2})  \right|^{2} \nonumber \\
  \hspace{-2cm}&+& \frac{1}{4 \pi^{2}} \left| \sum_{n_{1}=0}^{+\infty}\sum_{n_{2}=0}^{+\infty} q_{n_{1}} q_{n_{2}} C(n_{1},n_{2}+1,t) \exp(- i n_{1} \theta_{1}) \exp(- i n_{2} \theta_{2})  \right|^{2}.
 \end{eqnarray}
On the other hand, based on the Shannon's idea \cite{shannon} which is related to classical information theory and following the path of \cite{PB}, one may define the (Shannon) entropies associated with the number and phase probability distribution by the following relations:
\begin{eqnarray}\label{shannonent}
R_{n}(t) &=& - \sum_{n_{1}=0}^{+\infty} \sum_{n_{2}=0}^{+\infty}  \mathcal{P}_{n}(n_{1},n_{2}) \ln \mathcal{P}_{n}(n_{1},n_{2}), \nonumber \\
R_{\phi}(t) &=& - \int \limits_{\theta_{0}}^{\theta_{0}+2\pi}
d\theta_{1} \int \limits_{\theta_{0}}^{\theta_{0}+2\pi} d\theta_{2}
\; \mathcal{P}_{\phi}(\theta_{1},\theta_{2}) \ln
\mathcal{P}_{\phi}(\theta_{1},\theta_{2}),
\end{eqnarray}
where $\mathcal{P}_{n}(n_{1},n_{2})=\langle n_{1},n_{2}| \hat{\rho}_{F} |n_{1},n_{2} \rangle$ and $\hat{\rho}_{F}$ is the reduced density operator of the field. It has been shown that the sum of the entropies in equation (\ref{shannonent}) determines the lower bound of the entropy by the relation
$R_{n}+R_{\phi} \geq \ln 2\pi$. Considering this inequality, we suggest two quantities as follows:
\begin{eqnarray}\label{entsq}
E_{n}(t) = \frac{1}{\sqrt{2\pi}} \exp(R_{n}(t))-1,        \;\;\;\;\;\;\;\;\;\;            E_{\phi}(t) = \frac{1}{\sqrt{2\pi}} \exp(R_{\phi}(t))-1.
\end{eqnarray}
Whenever $-1<E_{n(\phi)}(t)<0$, the number (phase) component of the field entropy is squeezed. It is worthwhile to declare that the negative values of $E_{n(\phi)}(t)$ is another expression of the fact that $R_{n(\phi)}(t)$ is below its minimum value.\\
Figure \ref{Fnpeur} shows the time evolution of the phase entropic squeezing as a function of the scaled time $\tau$ for chosen parameters as considered in figure \ref{Finversion}. Also, the upside (downside) plots concern with the similar (different) amount of atom-field couplings.
From figure \ref{Fnpeur}(a) which corresponds to the resonance condition and no Kerr medium, it is observed that the phase entropic squeezing is appeared until the scaled time gets nearly the value $\tau = 11$.
Looking at figure \ref{Fnpeur}(b) indicates that in the presence of detuning parameters, the phase entropic squeezing occurs at all time. Figure \ref{Fnpeur}(c) which focuses on the influences of Kerr medium shows that, the phase entropic squeezing is pulled down (except in very small interval of time) due to the Kerr medium.
Considering both plots of figures \ref{Fnpeur}(c) and \ref{Fnpeur}(d) (studying simultaneously all considered nonlinearities) shows that they are qualitatively the same. \\
So, it may be inferred that detuning parameters have the outstanding role in exhibiting the phase entropic squeezing at all times and strengthen the amount of this nonclassicality feature. Also, fast oscillations can be regarded as a consequence of the presence of the detuning parameters. In addition, it is found that Kerr nonlinearity ruins the phase entropic squeezing.
%
 %===================================================================================================================================
 \subsection{Position-momentum entropic uncertainty relation and entropy squeezing}
 %===================================================================================================================================
Considering $\hat{x}$ and $\hat{p}$ as two observables (quadrature operators of the radiation field), Heisenberg uncertainty principle \cite{heisenberg} reads as $\Delta \hat{x} \Delta \hat{p} \geq 1/2$, where $\Delta \hat{x}$ ($\Delta \hat{p}$) shows the variance of the Hermitian operator $\hat{x}$ ($\hat{p}$). However, it should be noticed that the variance, which is used to define some quantum-mechanical effects such as quadrature squeezing of quantum
fluctuations, is not the only measure of quantum uncertainty, and sometimes the `entropy' may be preferred instead of the `variance'. Or{\l}owski \cite{Orlowski} has shown that, besides the fact that, the entropic uncertainty relation is stronger than the standard uncertainty relation, the entropy (of the single observable) as well as the variance can be utilized as a measure of squeezing of quantum fluctuations. \\
Taking into account the above explanations and using the Shannon's idea \cite{shannon}, one may introduce the entropies of position and momentum $E_{x} = - \int P(x) \ln P(x) dx$ and $E_{p} = - \int P(p) \ln P(p) dp$, where $P(x)$ and $P(p)$ are defined as $P(x) = \langle x|\hat{\rho}_{f}|x\rangle $ and $P(p) = \langle
p|\hat{\rho}_{f}|p\rangle $, respectively. In this case, the sum of the above-mentioned position and momentum entropies leads to the interesting inequality $E_{x}+ E_{p} \geq 1 + \ln \pi$ \cite{Orlowski}. This is often called the position-momentum entropic uncertainty relation that has been proven by Beckner \cite{proof-EUR} for the first time. It is valuable to notice that the entropies of position and momentum corresponding to the standard coherent state (and the vacuum state) are equal to each other, that is, $E_{x}=E_{p}=\frac{1}{2}(1+\ln \pi)$ \cite{Orlowski}. Using the latter inequality, one can supply an alternative mathematical formulation of the uncertainty principle by the inequality $\delta x \delta p \geq \pi e$ \cite{honarasaphase,ERspringer} where $\delta x $ and $\delta p $ corresponding to the exponential of Shannon entropies associated with the probability distributions for $x$ and $p$ are given by
 \begin{eqnarray}\label{shen}
 \delta x &=& \exp\left(-\int \limits_{-\infty}^{+\infty}\langle x|\hat{\rho}_{F}|x\rangle \ln \langle x|\hat{\rho}_{F}|x\rangle dx \right), \nonumber \\
 \delta p &=& \exp\left(-\int \limits_{-\infty}^{+\infty}\langle p|\hat{\rho}_{F}|p\rangle \ln \langle p|\hat{\rho}_{F}|p\rangle dp \right).
 \end{eqnarray}
In these relations, $\hat{\rho}_{F}$ is the reduced density operator of the field and its elements may be found as follows
 \begin{eqnarray}\label{xrx}
 \langle x|\hat{\rho}_{F}|x\rangle &=& \left| \sum_{n_{1}=0}^{+\infty} \sum_{n_{2}=0}^{+\infty} q_{n_{1}} q_{n_{2}} A(n_{1},n_{2},t) \langle x|n_{2}\rangle\right|^{2}  \nonumber \\
 &+& \left| \sum_{n_{1}=0}^{+\infty} \sum_{n_{2}=0}^{+\infty} q_{n_{1}} q_{n_{2}} B(n_{1},n_{2}+1,t) \langle x|n_{2}+1\rangle\right|^{2} \nonumber \\
 &+& \left| \sum_{n_{1}=0}^{+\infty} \sum_{n_{2}=0}^{+\infty} q_{n_{1}} q_{n_{2}} C(n_{1},n_{2}+1,t) \langle x|n_{2}+1\rangle\right|^{2},
 \end{eqnarray}
in which
 \begin{eqnarray}\label{x-n}
 \langle x|m \rangle=\left[ \frac{\exp\left(-x^{2}\right)}{\sqrt{\pi}2^{m} m!} \right]^{1/2} H_{m}(x),
 \end{eqnarray}
with $H_{m}(x)$ as the Hermite polynomials. Also, as is seen from the relations (\ref{xrx}) and (\ref{x-n}), we restricted our calculation to the field mode 2. \\
Now, in order to analyze the entropy squeezing properties of the system, we introduce two normalized quantities
 \begin{eqnarray}\label{quan}
 E_{x}(t) = (\pi e)^{-1/2} \; \delta x -1,        \;\;\;\;\;\;\;\;\;\;            E_{p}(t) = (\pi e)^{-1/2} \; \delta p -1.
 \end{eqnarray}
When $-1<E_{x(p)}(t)<0$, the position (momentum) or $x (p)$ component of the field entropy is said to be squeezed.\\
Figure \ref{Fpmeur} describes the position entropic squeezing for different chosen parameters, as considered in figure \ref{Finversion} with similar (different) atom-field couplings relating to the left (right) plots.
Figure \ref{Fpmeur}(a) is plotted for the situation in which no nonlinearity exists. This plot shows that, the position entropic squeezing gets negative values at all times nearly its minimum value ($-1$).
Adding the effect of the detuning parameters (figure \ref{Fpmeur}(b)) leads to the maximally position entropic squeezing all the time.
Figure \ref{Fpmeur}(c) analyzes the effect of Kerr medium. It is observed from the left plot of this figure that, nonclassicality behavior is destroyed in the  considerable intervals of time. By changing the ratio of atom-field couplings ($\gamma$) in the right one, position entropic squeezing gets minimum value at all times.
In figure \ref{Fpmeur}(d), where all nonlinearities are considered, it is seen that the nonclassicality behavior is lost when the time goes up. \\
Finally, one can deduce that while the detuning parameters play an important role to strengthen this nonclassical behavior (negativity of position entropic squeezing), the existence of the Kerr medium destroys this nonclassicality indicator. Moreover, changing appropriately the ratio of atom-field couplings may enrich and strengthen the  position entropic squeezing.

%============================================================================================================
 \section{Summary and concluding remarks}\label{examples}
%============================================================================================================
In this paper, we have studied a $\Lambda$-type three-level atom interacting with two coupled quantized fields which are simultaneously injected to a bichromatic cavity enclosed by the centrosymmetric Kerr medium in the presence of detuning parameters. Next, considering suitably all existing interactions, such as field-field interaction in the form of parametric down conversion, we have used the canonical transformation to simplify our complicated model to the usual form of the JCM. After obtaining the explicit form of the state vector of the whole system analytically, the effects of Kerr medium and detuning parameters on some of the well-known nonclassicality features have been individually and simultaneously examined. To reach this goal, the dynamics of atomic population inversion and also the time evolution of two different types of entanglement measures consist of von Neumann entropy (to study the tripartite entanglement) and linear entropy or idempotency defect (to discuss the decoherence) have been numerically studied. In addition, by using the number-phase and position-momentum entropic uncertainty relations, two types of entropy squeezing are evaluated. The main results of the paper are listed in what follows. \\
\begin{itemize}
 %=======================================================================
  \item {\it Atomic population inversion:} It is found that the presence of the detuning parameters leads to more visibility of the atomic population inversion. Adding the Kerr effect implies the fact that the population inversion is ruined. In other words, no population inversion exists when the Kerr effect is considered. It is also worthwhile to mention that the typical collapse and revival exhibition as a nonclassical phenomenon can be observed when the nonlinear couplers are absent. \\

%=======================================================================
  \item {\it Quantum entanglement:} It is observed that, the effect of nonlinearities is sharply pulled down the entanglement between the atom and the fields. It should be mentioned that changing the ratio of atom-field couplings ($\gamma$) has remarkable role on the amount of the DEM. In detail, in the presence/absence of all nonlinearities the DEM is descended while in the presence of either detuning parameters or Kerr medium, the amount of entanglement is ascended. In addition, oscillatory behavior is a result of the presence of the detuning. \\

%=======================================================================
  \item {\it Coherence loss:} It is shown that the coherence loss is attenuated due to nonlinear couplers. Also, in the presence of Kerr effect, selecting appropriate values of coupling constants may improve the decoherence. In this case, entering the detuning parameters results in the reduction of the linear entropy. Looking deeply at the numerical results of linear entropy and comparing them with the ones of von Neumann entropy indicates that, even though the general behavior of both of them are similar, for our system together with the considered conditions the measure of von Neumann entropy is more sensitive than the linear entropy. \\

%=======================================================================
  \item {\it Number-phase entropic squeezing:} It is illustrated that, detuning parameters have an effective role in revealing the phase entropic squeezing at all times and strengthen the amount of the negativity of this nonclassicality sign. Also, fast oscillatory can be thought of as a consequence of the detuning parameters. In addition, it is inferred that, the phase entropic squeezing is ruined when the Kerr nonlinearity is considered. \\

%=======================================================================
  \item {\it Position-momentum entropic squeezing:} As a consequence of the presence of the detuning parameters, position entropic squeezing is observed at all times, i.e., the time evolution of the related parameter remains always in the negative region and gets the constant value $-1$. But, this nonclassicality indicator is destroyed when the Kerr effect is present. Moreover, changing the ratio of atom-field couplings may enhance and strengthen the position entropic squeezing. \\

\end{itemize}
%=======================================================================
Finally, we would like to emphasize the generality of our model in the sense that this study can be performed by considering different configurations of three-level atom ($\Xi$- and $V$-type configurations) and/or different preparations for the initial states of the atom and/or the fields.
%=============================================================================================================
%
\vspace{1.00cm}
\\
 \textbf{\large{Acknowledgment}} \hspace{0.25cm} The authors would like to thank the referees for their valuable comments which improved the contents of the paper. One of the authors (MJF) wishes to dedicate this work to Hajar Mahmoudi for her kindness, selflessness and endless support which will always be remembered, and to the memory of Mohammad Haddad whose words of encouragement ring in his ears. Also, he is grateful to Ali Shakiba and Hamid Reza Baghshahi for useful discussion.
%=============================================================================================================
%=============================================================================================================
 %\newpage

 \end{document}